\documentclass[a4paper]{article}

\usepackage[english]{babel}
\usepackage[utf8x]{inputenc}
\usepackage[T1]{fontenc}
\usepackage{mathtools}
\usepackage{comment} 
\usepackage{tikz}
\newcommand*\circled[1]{\tikz[baseline=(char.base)]{
            \node[shape=circle,draw,inner sep=2pt] (char) {#1};}}
\usepackage{cancel}
\usepackage[a4paper,top=3cm,bottom=2cm,left=3cm,right=3cm,marginparwidth=1.75cm]{geometry}
\usepackage{amsmath}
\usepackage{graphicx}
\usepackage[colorinlistoftodos]{todonotes}
\usepackage[colorlinks=true, allcolors=blue]{hyperref}

\title{Heuristics for publishing dynamic content as structured data with schema.org}
\author{Elias K\"arle, Dieter Fensel
\\Semantic Technology Institute, University of Innsbruck
\\\{elias.kaerle, dieter.fensel\}@sti2.at}

\begin{document}
\maketitle

\begin{abstract}
Publishing fast changing dynamic data as open data on the web in a scalable manner is not trivial. So far the only approaches describe publishing as much data as possible, which then leads to problems, like server capacity overload, network latency or unwanted knowledge disclosure. With this paper we show ways how to publish dynamic data in a scalable, meaningful manner by applying context-dependent publication heuristics. The outcome shows that the application of the right publication heuristics in the right domain can improve the publication performance significantly. Good knowledge about the domain help choosing the right publication heuristic and hence lead to very good publication results.
\end{abstract}

\section{Introduction}
"Voice is the new search" they say and devices like Amazon's Echo, Apple's Siri, Google's Allo or Microsoft's Cortana, are about to change the way we search for information or consume content on the web. Those devices, called Intelligent Personal Assistants (IPA), alongside chatbots are summarized under the name "intelligent agents". Extracting useful information from websites to feed those agents is hard, since the data mostly is not structured and hence not implicitly machine read- and understandable. In 2011 the four leading search engine providers, Bing, Google, Yahoo! and Yandex introduced schema.org\footnote{http://schema.org}, a de-facto standard for structuring content on the web and hence make it machine read- and interpretable. Besides the beneficial SEO\footnote{Search Engine Optimization} effects of enriching Web content with schema.org, annotated data can also be consumed by third party software and accessed as a database-like knowledge source. So through annotation with schema.org websites can become valuable data sources for automated agents.

But the annotation of dynamic data is, due to its fast changing nature, a non trivial task. Dynamic data is, for example, availabilities or prices of hotel rooms. Whenever a room is booked, its availability changes and the data set is outdated. Prices of products in an online store, if their availability and price is subject to fast change, is considered as dynamic data. And also information about events and tickets are considered dynamic data since prices, availabilities and even the event itself changes over time. Manual annotation methods are therefore mostly no option because they don not scale. For automatic annotations several problems arise, like the computational power to calculate annotations on the server side or network latency due to big annotation overhead on a website. But also issues like content conformity with search engine guidelines and involuntary contingent disclosure are things that should be considered when publishing dynamic data in the form of schema.org annotations to the web.

With this paper the authors present heuristics to publish dynamic data of products and services on the provider's website as structured data. Different approaches are presented to overcome the problems that arise when trying to publish a huge amount of fast changing data. The first approach tackles the problem by only publishing a small amount of abstractions of the actual data. With an additional reference to a web service that can look up the concrete manifestations the abstract data gets elevated to being Semantic Web services. The second approach publishes one concrete instance of the data and allows the assumption that other variations are available too. Both approaches come with advantages and disadvantages which will be described below.

The reminder of this paper is structured as follows: Section \ref{sec:problem} gives a problem statement and derives the research questions. Section \ref{sec:ph-motivation} discusses related work and states the motivation for the work. Section \ref{sec:approach} shows the solution approach to the problem and answers the research questions. Section \ref{sec:evaluation} evaluates the approach and mentions limitations and Section \ref{sec:ph-conclusion} concludes the paper and gives an outlook on further challenges and applications.

\section{Problem Statement \& Research Question}
\label{sec:problem}
Publishing dynamic data\footnote{\textit{Dynamic data} is frequently changing information, like prices or availabilities of products. \textit{Active data} is information about interaction interfaces, e.g. Web Services.} as open data on the web seems like a rather trivial task as soon as the vocabulary and the publication techniques are in place. But if the the data to be published becomes more complex and transient, several problems occur.

\subsection{General problem}
In generic terms, the publication of dynamic data can be seen as the materialization of an algorithm. If we assume a simple multiplier algorithm, taking two integer numbers as input and performing a multiplication, the input-output table\footnote{https://en.wikipedia.org/wiki/Input-output\_model} looks as depicted in Table \ref{tab:io}.
\begin{table}[ht]
	\centering
	\begin{tabular}{c|c c c c c}
		$\cdot$ & 1 & 2 & 3 & \ldots & 1000 \\ \hline
        1 & 1 & 2 & 3 & \ldots & 1000 \\
        2 & 2 & 4 & 6 & \ldots & 2000 \\
        3 & 3 & 6 & 9 & \ldots & 3000 \\
        $\vdots$ & $\vdots$ & $\vdots$  & $\vdots$  & $\ddots$ & $\vdots$ \\
        1000 & 1000 & 2000 & 3000 & \ldots & 1000000
	\end{tabular}
	\caption{\label{tab:io}Input-output table for a multiplier with integer numbers from 1 to 1000 as an input.}
\end{table}
Even for this simple algorithm the input-output table contains one million results. This general problem occurs in various manifestations and was described for example in the context of Web service discovery in \cite{fensel2005wrong} or earlier in the context of expert systems in \cite{clancey1985heuristic}.

\subsection{Specific problem}
In more specific terms, when dynamic data represents products or services as output of complex algorithms based on several inputs the resulting input-output table explodes. For the ease of readability we will use the term \textit{product} synonymously for \textit{product}, \textit{service}, or in general, \textit{dynamic data} from now on. Further, the notation of the annotation vocabulary of our choice, \textit{schema.org}, will be abbreviated to \textit{sdo}.

The annotation of a single product with sdo to be published on a website as open date is straight forward. We use sdo/Product, add the product's name, description, an image and a sdo/Offer to publish a price, an availability and sdo/areaServed to specify from where the product can be bought or to where it is shipped. Various of those annotations can be found on eBay\footnote{https://www.ebay.com/} or other websites. We call this a zero-dimensional product, because no other variations of the same product are described on the same site.

If a product has variations, say, it is available in different colors, it becomes one-dimensional. This still is not a serious issue, because still, one website can specify products in several variations, and sdo can handle that by simply annotating a list of products, mentioning the different colors in the description or providing several offers in the sdo/offers property - maybe even with different prices and different availabilities. The same applies if we add a second dimension, where one product is available in several colors and maybe even sizes (thinking of cloths).

Multidimensional products make this annotation method unfeasible. If we imagine a hotel offering rooms, we talk about a multidimensional product. A hotel has several rooms (0D) with different numbers of beds inside (1D) and different amenity features (size of bathroom, TV, fridge, \ldots) (2D++). But the real multidimensional complexity uncovers when we consider that every room can be booked several times, if we add a temporal dimension, and can be booked either for one or for several days if we consider the dimension of the occupancy duration. Even if the same room, with a fixed number of beds and the same amenities, at the same time, for the same occupancy duration is booked by one or more people (double occupancy) the price changes and hence makes it technically a different product. Let alone the fact that different customers get different prices depending on the booking season, their booking time, their occupancy duration and possibly their loyalty status.

To show the complexity on some numbers we approximate a number of product variations $(P_v)$ for a small hotel. We assume ten rooms $(r=10)$ with two beds each (hence two occupation variations $(o=2)$), an average stay duration of seven days and an in-advance booking possibility of six months (30 days per month for six months and an average stay of seven days makes it about 25 bookings per room $(t=25)$) and two catering variations with \textit{breakfast} and \textit{half board} $(c=2)$:
\begin{equation} \label{eq:complexity}
P_v \coloneqq r \cdot o \cdot t \cdot c = 10 \cdot 2 \cdot 25 \cdot 2 = 1000
\end{equation}
If said hotel in that very simplified example, wants to annotate their booking web page, it has to create 1000 sdo/Offers. Now 1000 does not sound like a big number, but if we consider an annotation with 20 lines of code with 25 characters each (derived from calculations done on the work by \cite{akbar2017complete}) we have 500byte of data, times 1000 is half a megabyte data overhead on every single page load. That leads to new problems, which are discussed in the section below. For complex annotation and publication scenarios like the aforementioned, a publication methodology is needed which will be the subject of that paper.

\subsubsection{Infrastructural limitations}
From the problem mentioned above some critical infrastructural limitations arise.

\textbf{Aggregation latency vs. temporal accuracy:} For every single page call all possible product variations have to be calculated by a software running in the back-end of the website\footnote{For the aforementioned example either an internet booking engine (IBE), a property management system (PMS) or alike}. This leads to a heavy workload on the server, hence a page loading time latency and, in times of heavy website traffic, even a server crash. A countermeasure would be to have a ready-made set of annotations in place, with the crucial drawback of a very short-dated data accuracy. By the time one availability changes to "not available", the annotation data is already outdated. So this is, realistically, no feasible option. We refer to this problem as the \textit{latency vs. accuracy} dilemma.

\textbf{Network traffic limitation:} Considering the example in Equation \ref{eq:complexity} and assuming that each annotation consists of at least 20 lines of code with an average number of 25 characters we reach an annotation overhead of 500KB per page load. If we multiply the result of the example with a factor of ten, which is, due to its simplicity, very likely, every page load has to deliver 5MB of extra content, which leads to an unbearable loading delay. For a big product website, like a big hotel, or Amazon, the overhead would reach far more than several Megabytes. For this problem there is not even an out of the box workaround in place.

\textbf{Thought Experiment:} Let us develop the thought of having a sdo representation for every possible entity further. If we materialize every output an algorithm produces based on database entries to construct annotations, we could as well just save the materialized table, the list of annotations and replace the database completely. This materialization file (all possible annotations), depending on the type of information and on the number of entries, would probably be huge and the website loading overhead would go into gigabytes. This would make a delivery of the content as a website completely impossible and hence the annotations useless.

\subsubsection{Content conformity restriction}
Accuracy shortage appears as soon as the human readable website content does not match the web site's meta data. Back in the day, long before complex search algorithms, the search engines relied on meta tag descriptions of the content. This led to abuse by spammers who put way more, and even different, content into the meta data as the website actually contained to mislead search engines for their benefit. The search engine providers reacted by punishing websites who's meta data did not match the web site's content. This also applies for schema.org markup which is technically seen meta data. So if a web page would contain annotations of all product variations without a respective representation in the front end (which hardly would make sense), search engine providers would punish and rate down the whole website.

\subsubsection{Transparency}
Despite all the benefits of publishing available products and product variations as open data in real time, this might lead to some serious competitive issues. Having all products published as open data in real time means, that not only product aggregators and resellers can access and use the data, but also possible competitors. So the publication leads to an \textit{involuntary contingent disclosure} and transparency. In some businesses this might not be a problem, but, for example, in the hotel business a competing hotel could base their yield management on contingents available in hotels in their vicinity or hotels with similar target audiences.

The five problems mentioned above are summed up in Table \ref{tab:problems}.
\begin{table}[ht]
	\centering
	\begin{tabular}{l||c}
		\textbf{Problem occurrence} & \textbf{Problem manifestation} \\ \hline \hline
        Web server & \circled{1} aggregation latency vs. \circled{2} temporal accuracy \\
		Internet infrastructure & \circled{3} network traffic\\
        Web front-end & \circled{4} content conformity vs. \circled{5} involuntary contingent disclosure
	\end{tabular}
	\caption{\label{tab:problems}Summery of dynamic data publication problems.}
\end{table}

\subsection{Different problem perspectives}
We can differentiate between technical problems (1-3) and non-technical problems (4,5). Depending on the perspective from which we look at the problems they are either solvable easier (B2C perspective) or harder and require explicit publication heuristics (B2B). Both perspectives are described below.

\subsubsection{Business to Consumer (B2C) perspective}
This perspective has relatively little usage or relevance, because mostly direct consumer are not the target audience for annotations. Still, there exist use cases where client software extracts information from annotations on the fly. In the B2C perspective we assume a website which presents products to a consumer who visits directly over a web browser, or accesses the data over a third party API implementation (mobile app, intelligent personal assistant, chat bot or alike). In this case we can assume, that neither the web interface, nor the API presents all product variations at once, but in human consumable units. This is done by categorizing products over one or more dimensions and paginating the results. For the hotel example the user could only view the category of \textit{double rooms} for a certain time frame, hence two dimensions are already expanded. The remaining amount of products is manageable, and can be displayed in one website or, if it is still too much, be paginated over several result pages. Here problem \circled{1} is eliminated, because per result only a minor additional calculation has to be done for producing the annotations. Problem \circled{2} is eliminated, because the results are freshly generated for every search. Problem \circled{3} is also not existent due to the small number of results per page. Problem \circled{4} does not exist since every annotation correlates with one product in the user interface and problem \circled{5} is non existent because the transparency is already given by the web site's nature and hence the contingent disclosure is by choice.

\subsubsection{Business to Business (B2B) perspective}
Annotations are usually consumed by applications like search engines or aggregation platforms. If all products present on a website should for example be indexed by Google to be shown as rich results\footnote{https://developers.google.com/search/docs/guides/search-features} it is very important to have a proper publication solution which is not constrained by any of the aforementioned problems. If a website wants to have all its products included in an aggregation platform, like the idea mentioned in \cite{kaerle2017annotation}, it is crucial to provide meaningful publication methods that bypass the problems discussed.

\subsection{Research Questions}
\label{sec:rq}
The research questions derived from the above problem statement can hence be stated as follows:
\begin{enumerate}
\item Is there a way to publish dynamic data in a meaningful way overcoming problems such as latency, server overload or unwanted over-publications?
\item What are the trade offs of publication heuristics overcoming the problems discussed, against the bulk publication of the full data set as open data?
\item Is there a publication methodology that comes without any trade offs?
\end{enumerate}

\section{Motivation \& Related Work}
\label{sec:ph-motivation}
The challenge of publishing huge amounts of data, or also rediscovering data in huge data sets is not new. Some of the related works are closer to our open data on the web use case, some are more abstract but still related. The most interesting findings will be presented in this section.

In the work of \cite{omelayenko2001two} the authors present an integration approach to perform mappings between business' XML standards. In a two step process that data format is first mapped to RDF and then to the target format. In our case the target format would be schema.org in JSNO-LD serialization, but but the paper describes the mapping of all data available, which does make sense in the B2B case, but not for our use case, namely the publication on a website.

In \cite{klump2006data} the authors describe applying the principles of the 'Berlin Declaration' to data to have them available in a long term repository for reuse of the data by the scientific community. This thesis is not trying find methodologies for long living open data, but for data about transient information, like products and services and their availabilities.

In \cite{auer2009triplify} the authors present \textit{triplify}, a light weight software to publish data from relational databases on the web to different RDF serializations. This paper's emphasis lies not on the publication of a comprehensive open data representation of a relational database, but on the publication of a meaning- and useful subset of a database. As described above, a comprehensive product list, published on every website request, would probably over exceed every web servers capacity.

In \cite{khalili2012rdfa,Khalili2013} the authors present an editor extension to the well known TinyMCE editor as a WYSIWYM editor for authoring semantic annotations, called RDFaCE. This is a very convenient way to annotate static content but not appropriate for automatic content annotation. In \cite{karle2017semantify} the authors present a platform for a schema.org annotation creation and publication as well as an extension feature to automate annotation creation. In this approach the annotation files are stored inside a platform and then published to the respective website by either manual integration, with a JavaScript code that loads the annotations from the platform, or by using ready-made content management system plugins. This platform is feasible for static data or data produced by Cronjob based extensions but does not provide publication functionality for fast changing dynamic data.

The publication of the structured data found on the web will be enabled by semantically annotated web services. An early approach to that is described in \cite{ankolekar2002daml} where the authors present DAML-S, a DAML+OIL ontology for describing Web Services. In \cite{martin2004bringing} the authors describe how to annotate web services with OWL-S. In \cite{fensel2006enabling} the authors present WSMO, a web service modeling ontology. In \cite{kopecky2008hrests} the authors propose a machine-readable API description micro format they call hRESTS, which stands for HTML for RESTful Services. Built on top of hRESTS, the work in \cite{vitvar2008wsmo} describes WSMO-Lite, an ontology for annotating SOA web services. In \cite{roman2015wsmo} the authors even go a step farther and apply WSMO-Lite to RESTful Web services and present algorithms for Web service discovery. As opposed to the works presented above, this thesis focuses on the light weight web service annotation vocabulary of schema.org/Actions to enhance the data with services.

To attach a web service to the annotated objects published with the mentioned heuristics web service discovery is required. The work of \cite{rao2004survey} provides an overview of some approaches to compose Web services automatically and in \cite{milanovic2004current} four key issues for Web service composition are discussed. For this paper's approach the Web services will be composed and attached manually by the software publishing the annotations. But for future extensions of that work automatic web service discovery is an interesting enrichment. 

In \cite{dimou2018factors} the authors present an analysis of factors influencing linked data generation. The topic of materialization is discussed in a subsection where the authors distinguish between \textit{dumping} and \textit{on the fly materialization}. The problems of both methods regarding huge data sets are not discussed but maybe our idea can be inspiring for future works on that topic.

\section{Approach}
\label{sec:approach}
As described in Section \ref{sec:problem}, a publication of annotations of all possible product variations, or in other words a full materialization of a data set, is not feasible due to the \textit{latency vs. accuracy dilemma}, the \textit{network traffic limitation} and the \textit{content conformity restrictions} and might not be wanted by the product provider due to \textit{involuntary contingent disclosure}. This section presents different approaches to publish meaningful subsets of the full materialization.

\subsection{Abstraction}
\label{sec:abstraction}
One approach for meaningful publication of dynamic data is to not publish all concrete manifestations of the data but to generalize the data to an abstraction by removing its dimensions. This idea is based on the concept of \textit{heuristic classification} (\cite{clancey1985heuristic}) and was applied to \textit{Web service discovery} in the work of \cite{fensel2005wrong}.

In the abstraction process all detailed properties from the product are removed or grouped together in common superclasses. The product "hotel room for 2 people for 10 days starting at February 10" can for example be abstracted to "hotel room for more than one person in February". The most abstract representation of this is a the product without any properties, "hotel room". This abstraction makes sense if every dimension has a $length > 1$, so at least on variation. In the hotel example it makes no sense to abstract hotel room to accommodation because in this example are no other types of accommodations (tent, apartment, ...) possible.

To later match a user request and refine an abstract product to an actual product a web service to look up the concrete instance of that product is needed. This web service is connected to the abstract product. The web service takes all the parameters which were removed in the abstraction step as inputs and returns a book- or purchasable product. In schema.org (sdo) terms that means publishing an abstract offer and attaching a sdo/SerachAction to the offer. The input- and output parameters of the action define what is sent to and answered from the web service. Through the sdo/target property of the sdo/Action the consumer or client knows where to search for manifestations of that specific offer. The specified endpoint is the \textit{Web service}, used to retrieve the \textit{product (services)} offered by the \textit{service provider} and the sdo/SearchAction is the \textit{Web service description}. Hence, by the definition of \cite{fensel2005wrong}, adding a sdo/Action to the annotation makes the specified \textit{Web service} a \textit{Semantic Web service}, in fact, because of schema.org as the chosen annotation vocabulary, a very light-weight one. We call this conversion step, from an abstract product to a semantically fully described web service, elevation. To apply this publication method, the software taking care of publishing the data also has to take care of publishing the annotated abstractions of the data in combination with the semantic web service.

\textbf{Hotel example:} To demonstrate the abstraction process we assume a n-dimensional product, a hotel room. We assume rooms with single and double occupancy, and suites with single- double and triple occupancy. Additionally it offers \textit{breakfast} and \textit{half board} catering options. For simplicity reasons the hotel rooms can be only booked for one week or for two weeks and it is open all year. If we identify the dimensions we find (1) room type, (2) occupancy, (3) catering option, (4) stay duration and (5) booking time. The abstract products we infer from that are: a room offer, a room with an occupancy between one and three people, a room with unspecified catering options, a room with a stay duration of between one and two weeks and a room which can be booked any time of the year. Those abstract products can be further abstracted to one abstract offer 
and then the elevation step is applied to that offer. The publication of the hotel's room data to a website requires a software like an internet booking engine (IBE). This IBE typically has different endpoints where searches, bookings or purchases can be performed. To elevate the abstract product of our example to a Semantic Web service we attach a sdo/Action to the sdo/potentialAction property of the sdo/Offer or sdo/Product. This potential action describes all necessary inputs for searching for or booking a room in the hotel, like arrival date, number of guests, stay duration, catering options and others. It also specifies the output which is either a concrete product in the form of a sdo/Offer, or even a booking confirmation in the case of a booking or purchase.

\textbf{Trade off:}
This method comes with the trade off that the annotation on the website is never a real product but just an abstraction of it. The client/consumer has to query the attached web service every time he wants to learn it's manifestations and their availabilities. This might cause network latency or even extra server workload, which can affect the user's experience in a negative way.

\subsection{Specialization}
\label{sec:specialization}
Another approach for publishing dynamic data in a meaningful way is the \textit{specialization} approach. As opposed to the \textit{abstraction} approach we propose the publication of one concrete manifestations of products under the assumption that if variation \textit{a} of a product is available on a website then there might be also a variation \textit{b} available. This approach of course requires an elevation step as well, where the specific product annotation gets extended with a web service description (SWS). It also requires a-priori knowledge by the client/consumer regarding the potential availability of product variations. If the client has no knowledge about that possibility it might never trigger a search action for other product variations.

This approach has the advantage, that the publication step is very simple because the one product to be fully published can be selected arbitrarily. 
But it leaves also a lot more room for possible search dead-ends. Following the open-world assumption, every not specified product variation could possibly be available. Therefore a lot of search queries against the web server are potentially needed and the possibility of leading to no results is very high. To reduce the number of search queries and the number of requests with empty results some improvements could be implemented. The research fields of "learning by example" \cite{frasconi1995unified} and "case based reasoning" \cite{kolodner2014case} sound very promising but exceed the scope of that paper.

If we stick with the hotel example: Let us assume we publish an annotation for double bedroom between 12.2. and 19.2. for two people with \textit{half board} catering option. Since the annotations is available on the website this means, that the room offer is available. But due to the open-world assumption this means that possibly the same room is still available for the week after. And the same room offer is also available for the \textit{breakfast only} catering option. So we see, that we have one product which is assuredly available and immediately bookable, but all other possibilities are highly speculative and leave a lot of room for erroneous requests or not satisfying answers to web services.

\subsection{Type-level Materialization}
\label{sec:tlm}
As opposed to the afore mentioned approaches where either on abstract or one concrete product was published, this approach publishes more than one data set. The idea is to not abstract as rigorously as to reduce all the dimensions, but to publish one product per variation per dimension. For a t-shirt that comes in three colors, three sizes and three cutting styles a full materialization would result in 27 products ($3\cdot3\cdot3=3^3=27$)\footnote{Applying the \textit{Cartesian product}}. For the type-level materialization we only materialize one dimension at a time and one product for each variation and as a result only have nine product annotations ($3+3+3=9$).

\textbf{Elevation:} Of course this approach also needs an elevation step where the partly materialized products get a web service description. A n-dimensional product which has only one dimension materialized is not bookable and hence needs a sdo/SearchAction to let the consumer know how to search for bookable variations. Through this elevation step the partly materialized products again become Semantic web services.

\textbf{Trade off:} the trade off that approach comes with is again, that there is no concrete bookable product and that again one or several intermediate API calls are needed before a product can be finally booked or purchased.

\subsection{Selective instance-level Materialization}
\label{sec:silm}
In this approach materialization is not done on the type-level, but on the instance level. Since the instance level materialization is the root of all problems described in this paper, this approach is only selectively materializing some dimensions and ignoring others. For an n-dimensional product, not all dimensions have the same number of variations, the same length. There are shorter dimensions, with less variations and longer dimensions, with more variations. In the aforementioned t-shirt example, all three dimensions have only three variation and hence are short. A hotel room for example has short dimensions, like catering options, which hardly get longer than four variations (breakfast, half board, full board, all inclusive). But also very long dimensions like arrival date (every day of the year) or stay duration (1, 2, \ldots , n)\footnote{A maximum stay duration can not be known but for this case we use 30 days.}. The idea is to selectively materialize short dimensions and ignore long dimensions to keep the amount of materialized products manageable. The question arises what is a long dimension and how many dimensions can be cut to still make sense. This question is not to be answered generically because it is also dependent on the number of dimensions. For a one-dimensional product with a dimension length of 1000 it makes sens to materialize this dimension. If, in contrast, a ten-dimensional product has nine dimensions with a length of three and one dimension with the length of 25 then it makes sens to ignore that dimension. Generally speaking, the annotation overhead should not have any negative effect on the website's performance and hence has to be decided from use case to use case.

This approach is a mixture of the approaches mentioned in \ref{sec:abstraction} and \ref{sec:specialization} and seems to be the most promising approach. It provides the website with a reasonable amount of annotations for certain products to avoid technical problems and still does not come with other problems like content conformity violation or full contingent disclosure. But still the materialized products on the website are not concrete products and make the additional elevation step necessary.

\textbf{Elevation:} As mentioned in the other approaches, for retrieving a concrete product, the published products needs to be connected with a Web service description and hence become a Semantic Web service.

\textbf{Example:} To use the already familiar hotel example we assume a hotel with two room types (normal and comfort), two catering options (\textit{half board} and \textit{breakfast}), two occupancy options (single and double), the possibility to book in advance for a year and unrestricted stay durations. The short dimensions hereby are room type (length 2), catering option (length 2) and occupancy (length 2). The long dimensions are arrival data (length 365) and stay duration (not defined, but 1 day to 30 or more days is possible, hence length can be 30 or more). A full materialization results in 87600 products to be annotated (Equation \ref{eq:full-mat}),
\begin{equation} \label{eq:full-mat}
P(v_1) = 2 \cdot 2 \cdot 2 \cdot 365 \cdot 30 = 87600
\end{equation}
while a selective instance level materialization, where the dimensions \textit{arrival} and \textit{stay duration} are ignored, only results in 8 annotations (Equation \ref{eq:sil-mat}) with the references to the respective web services.
\begin{equation} \label{eq:sil-mat}
P(v_2) = 2 \cdot 2 \cdot 2 \cdot \cancel{365} \cdot \cancel{30}= 8
\end{equation}

\textbf{Trade off:}
As before, this method comes with the trade offs that by the time the semantic annotations are read by a client it is not clear if the product is currently available or not. So in general we assume, that the product is available if it has a representation. If a book- or purchase request is triggered, then the attached web service is requested to see if it is really available. To increase the hit ratio and lower the "product no longer available" errors, some statistical reasoning can be applied (statistics based on expert's experience, season, weather, locality's specialties (winter/summer sports, ...), ...), but this exceeds the scope of that paper.

\section{Evaluation \& Limitations}
\label{sec:evaluation}
This section compares the aforementioned approaches and evaluates them, based on five implementations, qualitatively and quantitatively. Furthermore the research questions will be answered and the limitations each approach comes with are discussed.

\subsection{Evaluation}
To evaluate and compare the publication heuristics we consider a fictional hotel with the following properties: it has 10 rooms of two types (\textit{normal, comfort}); it offers two catering options (\textit{breakfast, half board}); there are two occupancy options (\textit{single, double}); the maximum stay duration is limited to \textit{30} days; and a booking can be up to \textit{365} days in advance. As calculated in Equation \ref{eq:full-mat} this results into 87.600 different combinations if fully materialized.

To test the publication heuristics we set up a demo website describing said hotel\footnote{http://bache.rotes-wildschwein.at/}. The static data was annotated manually with schema.org and is the same for every test scenario. We simulate booking functionality by working with a demo account we got from the internet booking engine (IBE) company Feratel\footnote{feratel.at}.
The software does not use semantic annotations whatsoever, so we utilize a wrapper, that translates sdo/Actions to API calls and API responses to sdo/Things according to the approach described in~\cite{simsek2018apiwrapping}. To simulate that the demo website annotates their dynamic data, the offers, prices and availabilities, we made a script that requests the IBE's API. For every one of the four heuristics, \textit{Abstraction}, \textit{Specialization}, \textit{Type-level Materialization} and \textit{Selective instance-level Materialization}, the script prepares annotated data sets accordingly. Then, the results are displayed on individual web pages against which we run the tests. Additionally we also prepared a data set and a website with a full materialization as a fifth test scenario.

\subsubsection{Quantitative Evaluation}
\begin{figure}
\centering
\includegraphics[width=0.99\textwidth]{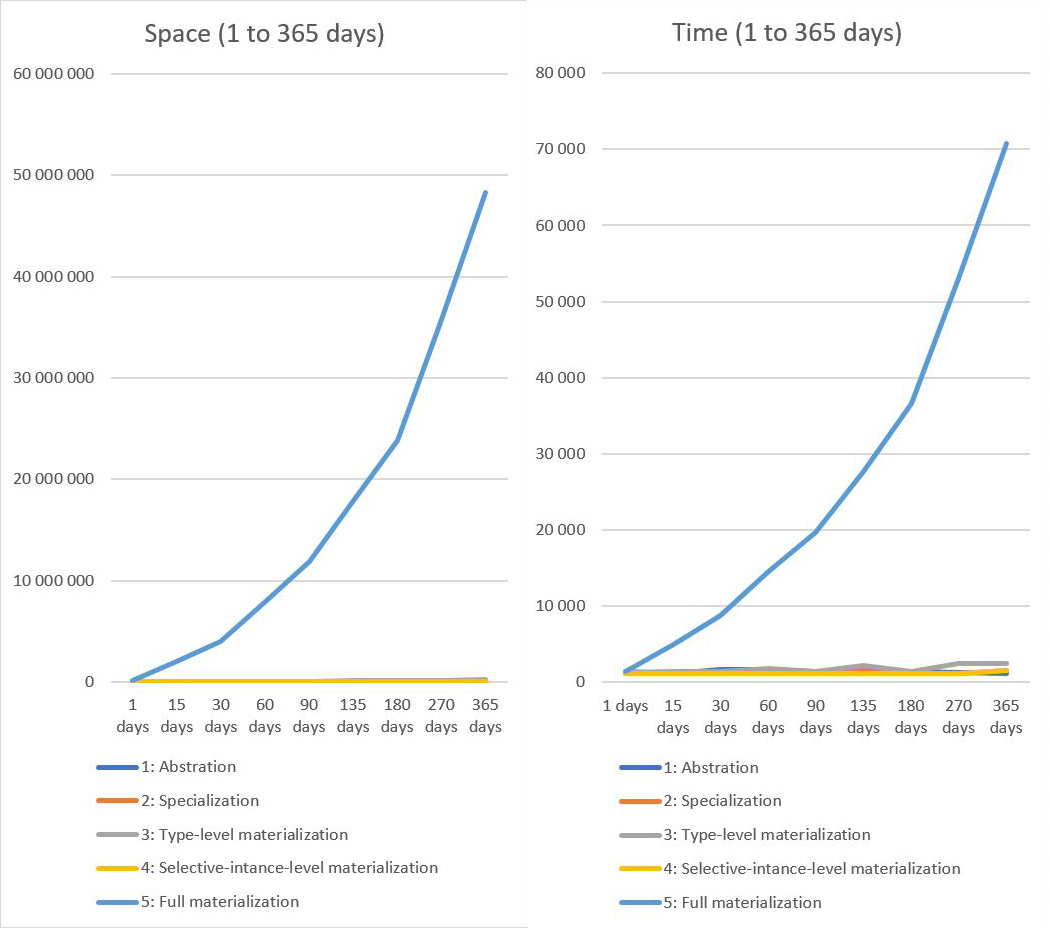}
\caption{\label{fig:space-time}\textbf{Left:} the space annotations consume within the demo website, by the four publication heuristics and the full materialization. The unit of the y-axis is Bytes. \textbf{Right:} The rendering times of the websites of the four publication heuristics and the full materialization. The unit of the y-axis is milliseconds.}
\end{figure}

\begin{figure}
\centering
\includegraphics[width=0.99\textwidth]{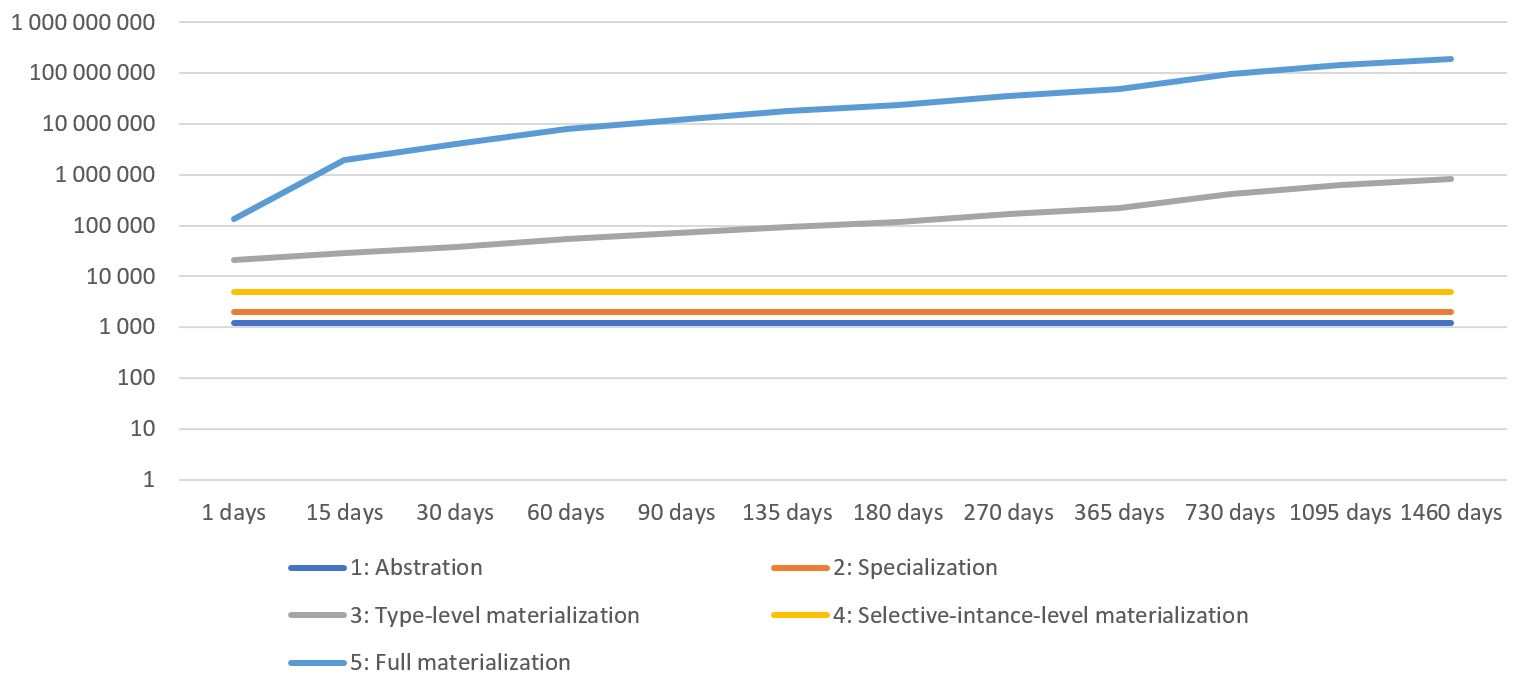}
\caption{\label{fig:space-log}The space annotations consume within the demo website, by the four publication heuristics and the full materialization on a logarithmic scale. The unit of the y-axis is Bytes.}
\end{figure}

\begin{figure}
\centering
\includegraphics[width=0.99\textwidth]{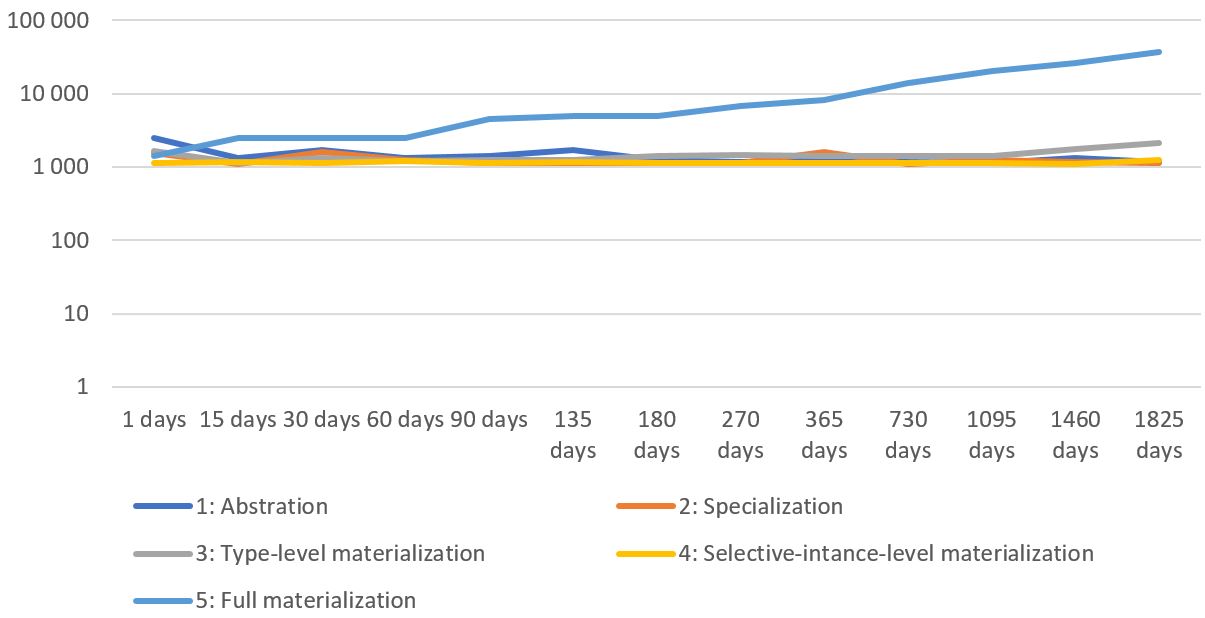}
\caption{\label{fig:time-log}The rendering times of the websites of the four publication heuristics and the full materialization on a logarithmic scale. The unit of the y-axis is milliseconds.}
\end{figure}

As described in Section \ref{sec:problem}, problems \circled{1} - \circled{3}, \textit{aggregation latency}, \textit{temporal accuracy} and \textit{network traffic}, concern the size of the generated and transmitted annotations and, related to that, the page loading latency. Therefore, to evaluate the different heuristics quantitatively, we measured the size of the web page where the respective publication heuristic was applied and the time it took to load the web page. To get a feeling when it starts to make sense to use publication heuristics instead of a full materialization we made the evaluation with a different number of variations. As described, the test case has five dimensions with dimension lengths of 2 (room types), 2 (catering options), 2 (occupancy options), 30 (max. stay duration) and 365 (in advance booking). To simulate a different number of product variations we made ten test runs where the last dimension, the number of days a booking can be made in advance, was flexible: starting from one, up to 1825, which would be five years\footnote{Of course, no one books five years in advance. But this example should show that for multi-dimensional products, long dimensions can be a problem.}. So the number of product variations, and hence annotations, varies from 240 up to 438.000. What that means for the extra space the annotations consume within a web page and the page's loading time can be seen in Figures \ref{fig:space-time} to \ref{fig:time-log}.

As can be seen in Figure \ref{fig:space-time} (left), the space consumed by annotations of product variations grows exponentially, with the factor $n$ where $n=$\textit{days in advance booking}. For an in-advance-booking of one year, the annotations for the demo hotel make already more than 50MB in a full materialization.

Figure \ref{fig:space-log} shows that publication heuristic 3 (Type-level materializaiton also grows to rather big annotations with summand $n$ where $n=$\textit{days in advance booking}. The other three publication heuristics do not grow with the increas of the in-advance booking option and hence lead to good results regarding the annotation size.

To achieve good measurements for the page loading time we repeated the experiment several times and used the mean time measurement in our evaluation. Figure \ref{fig:space-time} (right) shows, that only the full materialization really affects the page loading time and that all the publication heuristics lead to significantly better results.

In Figure \ref{fig:time-log} we see, that the page loading time is always around one second, but grows out of the reasonable quickly when applying full materialization. Even though all four publication heuristics have, under the tested circumstances, very good page loading times the type level materialization grows, compared to the other three, too fast and should be avoided for big data sets. When a data set is considered "big" and what are "long" dimensions, as mentioned in publication heuristic 4, is subject of future work and will be described at the end of that paper.

\subsubsection{Qualitative Evaluation}
As described in Section \ref{sec:problem}, problems \circled{4} and \circled{5} concern the \textit{content conformity} and the \textit{involuntary contingent disclosure}.

\textit{Content conformity} describes the fact that the content of the annotations should match the content of the web front end. In publication heuristic one and two it is very easy to satisfy content conformity. The web front end only has to specify on abstract product or one specific product variation. For publication heuristic three and four it is relatively easy to satisfy content conformity. The front end simply has to show that all the different options are possible and hence legit to be held in the annotation.

\textit{Involuntary contingent disclosure} describes the problem, that, trough extensive annotation of products, the competitor may can derive the contingent of a, for example, hotel and hence can benefit competitively from that knowledge. No publication scenario solves that completely but all four hide the full contingent reasonably. Due to the fact that all four heuristics add annotated web services to the product's annotation one can theoretically always retrieve all contingents. But this does not work out of the box, require still good knowledge of the underlying API and can not be achieved without considerable coding effort.

\subsection{Answering the research questions}
In the following section, the three research questions, asked in \ref{sec:rq} will be answered and justified.

1: \textit{Is there a way to publish dynamic data in a meaningful way overcoming problems such as latency, server overload or unwanted over-publications?} Yes. In this paper we presented four publication heuristics which, depending on the use case, give a guidance in publishing dynamic data.

2. \textit{What are the trade offs of publication heuristics overcoming the problems discussed, against the bulk publication of the full data set as open data?} The only trade off of using certain heuristics over a full materialization is, that not all concrete manifestations are published. But this, as shown, enables publication of dynamic data in the first place. So there is more trade off when trying to publish full materializations (namely a publication is not possible) than when applying the mentioned publication heuristics.

3. \textit{Is there a publication methodology that comes without any trade offs?} No. Every heuristic comes with a trade off. Publication heuristics 1 and 2 are the most easy to implement and the smallest in annotation space but offer the least information, heuristics 3 and 4 offer a lot of data but have no concrete manifestations.

\subsection{Limitations} It is obvious that the proposed solutions come with limitations, otherwise no heuristics would be necessary. During our evaluation we could identify the following still unsolved challenges.

The main limitation is that almost no concrete manifestations are published. Depending on the applied heuristic the consumer of the data has to know how to retrieve the concrete manifestations. This is, of course, described implicitly trough the published sdo/Actions, but still at least one additional step has to be performed to reach concrete manifestations of data.  Another limitation is, that the implementation of the heuristics can be tricky when it comes to deciding which heuristic is used and how, for example in \textit{selective instance-level materialization}, the "long" and "short" dimensions are chosen. To overcome that limitation further investigation will be part of the future work.

\section{Conclusion \& Future Work}
\label{sec:ph-conclusion}
With this paper we were describing the work done to find meaningful ways to publish dynamic, fast changing data on the web in a machine read- and interpretable way. Since a full manifestation of an algorithm is in the most cases not feasible and sometimes not even possible we tried to work around that issue. We presented four different approaches that publish either abstractions or single examples of the data. In both cases it was necessary to attach web service descriptions which point to APIs. In the abstraction case the description pointed to the service to retrieve the real manifestation of the data. In the case of a single published instance the web service points to a search interface to find other manifestations of the data. The evaluation has shown, that no clear winner of the four approaches could be nominated and that the application of publication heuristics is highly use case dependent. What the evaluation also showed is, that every approach, in every use case is better than a full manifestation of an algorithm.

Our future work goes into several directions. One thing that is going to be investigated is the question that was already asked in Section \ref{sec:silm}, what are long dimensions and when do long dimensions become a problem. Therefore we are going to take a look into the literature and define metrics when and under what circumstances a dimension is considered "long" and needs to be ignored for heuristics like \textit{selective instance level materialization}.

Of course the for heuristics presented are not the answer to everything. So we are also working on refinement and improving the presented heuristics and maybe come up with some new, even better, ideas.

Abstracting a dimension always comes with information loss which is, in our case, compensated by the introduction of a Semantic Web Service where the missing information can be looked up. But for more generic cases we are also taking a look in the field of "data compression with information loss" where we might combine our approach with well known data compression ideas like for example in \cite{burrows1994block}.

As practical future work we are planning to apply the learned to real life scenarios. We are already in talks with IBE providers we are working with where we want to help them implement the publication heuristics in their software. This would be a great opportunity for a long term real life analysis and insight into the performance of our proposed solutions.

We are also pursuing an extension of the semantic annotation distribution platform \textit{semantify.it}\cite{karle2017semantify}. Instead of publishing all the annotations of a user on his website with the semantify.it technology it would be thinkable to give the user the possibility to only publish certain subsets of his data based on the proposed publication heuristics.

To validate the semantically annotated dynamic data on a website, it would be interesting to extend the work described in \cite{simsek2017domain}. Instead of validating only based on the mentioned domains specifications (DS), it is also thinkable to validate towards our publication heuristics to avoid the issues mentioned in this paper.

A last outlook or potential future work concerns the annotation vocabulary. In \cite{karle2017extending} the authors describe the extension of the schema.org vocabulary to annotate dynamic data for the accommodation business. Other verticals, like the service industry, still lack such vocabulary, which would be an interesting use case for future research.

\section*{Acknowledgements}
The authors would like to thank the Online Communications working group (OC)\footnote{http://sti2.at} for their active discussions and input during the OC meetings and the highly committed \textit{semantify.it}\footnote{https://semantify.it} development team for their implementation efforts. Furthermore the authors want to thank \ldots .

\bibliographystyle{apalike}
\bibliography{bib}

\begin{thebibliography}{}

\bibitem[Akbar et~al., 2017]{akbar2017complete}
Akbar, Z., K{\"a}rle, E., Panasiuk, O., {\c{S}}im{\c{s}}ek, U., Toma, I., and
  Fensel, D. (2017).
\newblock Complete semantics to empower touristic service providers.
\newblock In {\em OTM Confederated International Conferences" On the Move to
  Meaningful Internet Systems"}, pages 353--370. Springer.

\bibitem[Ankolekar et~al., 2002]{ankolekar2002daml}
Ankolekar, A., Burstein, M., Hobbs, J., Lassila, O., Martin, D., McDermott, D.,
  McIlraith, S., Narayanan, S., Paolucci, M., Payne, T., et~al. (2002).
\newblock Daml-s: Web service description for the semantic web.
\newblock {\em The Semantic Web—ISWC 2002}, pages 348--363.

\bibitem[Auer et~al., 2009]{auer2009triplify}
Auer, S., Dietzold, S., Lehmann, J., Hellmann, S., and Aumueller, D. (2009).
\newblock Triplify: light-weight linked data publication from relational
  databases.
\newblock In {\em Proceedings of the 18th international conference on World
  wide web}, pages 621--630. ACM.

\bibitem[Burrows and Wheeler, 1994]{burrows1994block}
Burrows, M. and Wheeler, D.~J. (1994).
\newblock A block-sorting lossless data compression algorithm.

\bibitem[Clancey, 1985]{clancey1985heuristic}
Clancey, W.~J. (1985).
\newblock Heuristic classification.
\newblock {\em Artificial intelligence}, 27(3):289--350.

\bibitem[Dimou et~al., 2018]{dimou2018factors}
Dimou, A., Heyvaert, P., De~Meester, B., and Verborgh, R. (2018).
\newblock What factors influence the design of a linked data generation
  algorithm?
\newblock In {\em LDOW2018 workshop, part of WWW2018, the International World
  Wide Web Conference}, pages 1--6.

\bibitem[Fensel et~al., 2005]{fensel2005wrong}
Fensel, D., Keller, U., and Polleres, A. (2005).
\newblock What is wrong with web services discovery?
\newblock In {\em Position Paper for the Workshop on Frameworks for Semantics
  in Web Services}, Innsbruck, Austria.

\bibitem[Fensel et~al., 2006]{fensel2006enabling}
Fensel, D., Lausen, H., Polleres, A., De~Bruijn, J., Stollberg, M., Roman, D.,
  and Domingue, J. (2006).
\newblock {\em Enabling semantic web services: the web service modeling
  ontology}.
\newblock Springer Science \& Business Media.

\bibitem[Frasconi et~al., 1995]{frasconi1995unified}
Frasconi, P., Gori, M., Maggini, M., and Soda, G. (1995).
\newblock Unified integration of explicit knowledge and learning by example in
  recurrent networks.
\newblock {\em IEEE Transactions on Knowledge and Data Engineering},
  7(2):340--346.

\bibitem[K{\"{a}}rle and Fensel, 2017]{kaerle2017annotation}
K{\"{a}}rle, E. and Fensel, D. (2017).
\newblock Annotation-based automatic action processing.
\newblock In {\em Proceedings of the {ISWC} 2017 Posters {\&} Demonstrations
  and Industry Tracks co-located with 16th International Semantic Web
  Conference {(ISWC} 2017)}, Vienna, Austria.

\bibitem[K{\"a}rle et~al., 2017]{karle2017extending}
K{\"a}rle, E., Simsek, U., Akbar, Z., Hepp, M., and Fensel, D. (2017).
\newblock Extending the schema. org vocabulary for more expressive
  accommodation annotations.
\newblock In {\em Information and Communication Technologies in Tourism 2017},
  pages 31--41. Springer.

\bibitem[{K\"{a}rle} et~al., 2017]{karle2017semantify}
{K\"{a}rle}, E., {\c{S}im\c{s}ek}, U., and {Fensel}, D. (2017).
\newblock {semantify.it, a Platform for Creation, Publication and Distribution
  of Semantic Annotations}.
\newblock In {\em SEMAPRO 2017: The Eleventh International Conference on
  Advances in Semantic Processing}, pages 22--30. New York: Curran Associates,
  Inc.

\bibitem[Khalili and Auer, 2013]{Khalili2013}
Khalili, A. and Auer, S. (2013).
\newblock Wysiwym authoring of structured content based on schema. org.
\newblock In {\em WISE (2)}, pages 425--438.

\bibitem[Khalili et~al., 2012]{khalili2012rdfa}
Khalili, A., Auer, S., and Hladky, D. (2012).
\newblock The rdfa content editor-from wysiwyg to wysiwym.
\newblock In {\em Computer Software and Applications Conference (COMPSAC), 2012
  IEEE 36th Annual}, pages 531--540. IEEE.

\bibitem[Klump et~al., 2006]{klump2006data}
Klump, J., Bertelmann, R., Brase, J., Diepenbroek, M., Grobe, H., H{\"o}ck, H.,
  Lautenschlager, M., Schindler, U., Sens, I., and W{\"a}chter, J. (2006).
\newblock Data publication in the open access initiative.
\newblock {\em Data Science Journal}, 5:79--83.

\bibitem[Kolodner, 2014]{kolodner2014case}
Kolodner, J. (2014).
\newblock {\em Case-based reasoning}.
\newblock Morgan Kaufmann.

\bibitem[Kopeck{\`y} et~al., 2008]{kopecky2008hrests}
Kopeck{\`y}, J., Gomadam, K., and Vitvar, T. (2008).
\newblock hrests: An html microformat for describing restful web services.
\newblock In {\em Web Intelligence and Intelligent Agent Technology, 2008.
  WI-IAT'08. IEEE/WIC/ACM International Conference on}, volume~1, pages
  619--625. IEEE.

\bibitem[Martin et~al., 2004]{martin2004bringing}
Martin, D., Paolucci, M., McIlraith, S., Burnstein, M., McDermott, D.,
  McGuinness, D., Parsia, B., Payne, T.~R., Sabou, M., Solanki, M., et~al.
  (2004).
\newblock Bringing semantics to web services: The owl-s approach.

\bibitem[Milanovic and Malek, 2004]{milanovic2004current}
Milanovic, N. and Malek, M. (2004).
\newblock Current solutions for web service composition.
\newblock {\em IEEE Internet Computing}, 8(6):51--59.

\bibitem[Omelayenko and Fensel, 2001]{omelayenko2001two}
Omelayenko, B. and Fensel, D. (2001).
\newblock A two-layered integration approach for product information in b2b
  e-commerce.
\newblock In {\em International Conference on Electronic Commerce and Web
  Technologies}, pages 226--239. Springer.

\bibitem[Rao and Su, 2004]{rao2004survey}
Rao, J. and Su, X. (2004).
\newblock A survey of automated web service composition methods.
\newblock In {\em SWSWPC}, volume 3387, pages 43--54. Springer.

\bibitem[Roman et~al., 2015]{roman2015wsmo}
Roman, D., Kopeck{\`y}, J., Vitvar, T., Domingue, J., and Fensel, D. (2015).
\newblock Wsmo-lite and hrests: Lightweight semantic annotations for web
  services and restful apis.
\newblock {\em Web Semantics: Science, Services and Agents on the World Wide
  Web}, 31:39--58.

\bibitem[{\c{S}}im\c{s}ek et~al., 2018]{simsek2018apiwrapping}
{\c{S}}im\c{s}ek, U., K\"{a}rle, E., and Fensel, D. (2018).
\newblock Machine readable web apis with schema.org action annotations.
\newblock (To appear).

\bibitem[{\c{S}}im{\c{s}}ek et~al., 2017]{simsek2017domain}
{\c{S}}im{\c{s}}ek, U., K{\"a}rle, E., Holzknecht, O., and Fensel, D. (2017).
\newblock Domain specific semantic validation of schema. org annotations.
\newblock In {\em International Andrei Ershov Memorial Conference on
  Perspectives of System Informatics}, pages 417--429. Springer.

\bibitem[Vitvar et~al., 2008]{vitvar2008wsmo}
Vitvar, T., Kopeck{\`y}, J., Viskova, J., and Fensel, D. (2008).
\newblock Wsmo-lite annotations for web services.
\newblock In {\em European Semantic Web Conference}, pages 674--689. Springer.

\end{thebibliography}

\end{document}